\documentclass[showpacs,amsmath,amssymb,floatfix,prl,twocolumn]{revtex4}

\usepackage[dvips]{graphicx}
\usepackage{amssymb,amsfonts,amsmath}
\usepackage{graphicx,amsmath}

\begin{document}

\title{Enhancement of edge channel transport by a low frequency irradiation}
\author{A.D. Chepelianskii$^{(a,b)}$, J. Laidet$^{(a)}$, I. Farrer$^{(b)}$, H.E. Beere$^{(b)}$, D.A. Ritchie$^{(b)}$, H. Bouchiat$^{(a)}$\\
(a) LPS, Univ. Paris-Sud, CNRS, UMR 8502, F-91405, Orsay, France\\
(b) Cavendish Laboratory, University of Cambridge, J J Thomson Avenue, Cambridge CB3 OHE, UK}


\begin{abstract}
The magnetotransport properties of high mobility two dimensional electron gas have recently 
attracted a significant interest due to the discovery of microwave induced zero resistance states. 
Here we show experimentally that microwave irradiation with a photon energy
much smaller than the spacing between Landau levels can induce a strong decrease in the four terminal
resistance. We propose an interpretation of this effect based on the enhancement of the 
drift velocity of the skipping orbits along sample edges.
\end{abstract}

\maketitle

Transport under high frequency  microwave irradiation in high purity two dimensional electron gases (2DEG)
revealed many intriguing and unexpected phenomena of which 
microwave induced zero resistance (ZRS) states are probably the most striking manifestation.
As experiments in Refs. \cite{mani2002,zudov2003} show, microwave irradiation 
can lead to a complete disappearance of longitudinal resistance $R_{xx}$ for particular values 
of the ratio $j = \omega/\omega_c$ between the driving frequency $\omega$ and the cyclotron frequency $\omega_c$.
Until 2010 this dissipationless effect was only observed in GaAs heterostructures
of ultra high purity \cite{mani2002,zudov2003} or high densities \cite{bykov} . 
However the recent observation of ZRS for electrons on the liquid helium surface 
indicates that it is actually a generic effect that may appear in very different physical systems \cite{denis}.
Despite the important theoretical efforts that were made to understand this effect,
the physical origin of ZRS is still controversial.
Most widely accepted models \cite{ryzhii,girvin,polyakov,vavilov,khodas2} argue 
that microwave irradiation creates a negative resistance state which is unstable and gives 
rise to a zero resistance state through the formation of current domains.
However no conclusive experimental evidence has been provided in support of 
this scenario and some experimental features do not seem to be easily 
understood on the basis of the above picture.
In the ZRS regime resistance decreases exponentially 
with microwave power \cite{mani2004} and inverse temperature \cite{zudov2003},
instead of a direct switching to a non-dissipative state.
Also it was shown that zero resistance states are not affected by the sense of circular polarization
which questions mechanisms relying explicitly on transitions between Landau-levels \cite{smetpolar}.
Moreover ZRS disappear in Hall bars with a small channel size of a few microns, 
which indicates the importance of edge effects \cite{bykov2009}. 
These experimental properties highlight the difficulties encountered by the conventional theoretical descriptions of ZRS.
While a strong fraction of the theoretical community believes that these difficulties will be answered 
in a yet to-be developed theory of the zero resistance state, several recent proposals have attempted
to explain zero resistance states without appealing to an intermediate state of negative resistance 
\cite{toulouse,mixalkov}.

In this article we investigate the adiabatic limit $\omega \ll \omega_c$ where transitions between 
Landau-levels are excluded.  
We show experimentally that even in this case microwave irradiation can lead to a strong suppression 
of $R_{xx}$ in a wide range of magnetic fields. We then propose a semi-classical model 
that explains the observed effect through the enhancement of the drift velocity of trajectories 
skipping along sample edges, these results have a strong connection with the recent theory \cite{toulouse} 
which proposed that ZRS appears due to microwave stabilization of electron transport along sample edges. 
A decrease of $R_{xx}$ under irradiation at frequencies smaller than $\omega_c$ was already reported in \cite{dorozhkin}. 
However in this experiment the frequency of the exciting photons was around $20\;{\rm GHz}$ 
which is around an order of magnitude larger than the typical frequencies used in our experiments.
The ratio $\omega/\omega_c$ is therefore much smaller in the present experiments and we expect to 
be in a truly adiabatic limit even for low magnetic fields of $0.1\;{\rm Tesla}$.
In \cite{dorozhkin}, comparable values of the parameter $\omega/\omega_c$ could be reached 
only at much higher magnetic fields around $1\;{\rm Tesla}$ which 
corresponds to a different physical regime where Landau levels are well separated. 

\section{I. Experiment and effect of microwaves on group velocity} 

We have investigated magneto-transport under microwave irradiation in a $GaAs/Ga_{1-x}Al_x As$
2DEG with density $n_e \simeq 3.3\times10^{11}{\rm cm}^{-2}$, 
mobility $\mu \simeq 10^7\;{\rm cm^2/Vs}$ corresponding to transport time $\tau_{tr} \simeq 1.1\;{\rm ns}$.
The Hall bar with a 100 $\mu m$ wide channel was patterned using wet etching (see Fig.~1 inset).
A micro-bonding wire was positioned on the Hall bar chip, parallel to the current channel
at a distance of $100\;{\rm \mu m}$ from the nearest edge. 
One of the extremities of the wire was connected to a coaxial cable, which allowed to send microwave irradiation 
in a broad frequency range from $1$ GHz to $40$ GHz. The sample was cooled in a He$^{3}$ insert
to a temperature of around $500$ mK. 
We compared the effect of microwaves on the magnetoresistance at two different frequencies $f = 38.65\;{\rm GHz}$ and $f = 2.3\;{\rm GHz}$. 
As shown, on Fig.~1, the high frequency irradiation leads to microwave induced resistance oscillations (MIRO) 
similar to those reported in \cite{zudov2}. The magnetoresistance under irradiation 
is characterized by a series of peaks and dips as a function of magnetic field
whose position are determined by the ratio $j = \omega/\omega_c$ of the microwave frequency $\omega$
to the cyclotron frequency $\omega_c$. 
At higher microwave power these oscillations are expected to develop into ZRS,
however in our experiments this regime was out of reach due to the limited cooling power of the He$^{3}$ insert.
The magnetoresistance under high frequency microwave irradiation $f = 38.64\;{\rm GHz}$ contrasts sharply 
with the behavior under irradiation at $f = 2.3\;{\rm GHz}$. 
In the latter case the magnetoresistance does not exhibit oscillations anymore but presents a significant drop under irradiation 
in a large range of magnetic fields ($H \ge 0.05 \;{\rm Tesla}$). This drop can not 
be explained by an increase in electron temperature since resistance increases with temperature in the explored range of magnetic fields.

\begin{figure}
\centerline{\includegraphics[clip=true,width=8cm]{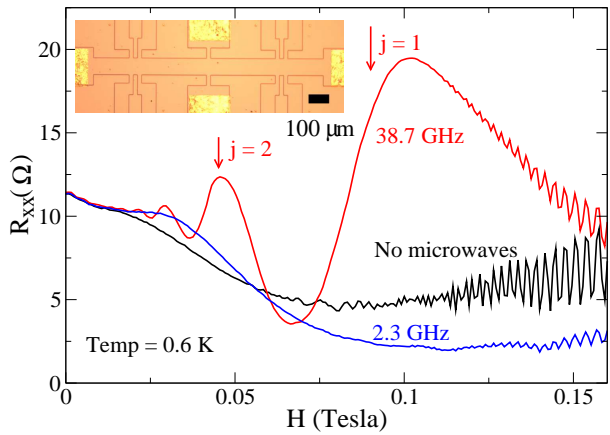}}
\caption{Magnetoresistance of a high mobility Hall bar (optical photograph of the sample is shown in the inset) 
in the absence of microwaves and under irradiation at $f = 38.7\;{\rm GHz}$ and $f = 2.3\;{\rm GHz}$.
The high frequency irradiation induces oscillations in the magnetoresistance (MIRO), 
whereas the low frequency driving leads to an homogeneous drop in $R_{xx}$ for $H > 0.1\;{\rm Tesla}$. 
}
\label{SemiFig1}
\end{figure}

It is difficult to explain these effects of low frequency irradiation with a purely bulk mechanism.
For a frequency $f = 2.3\;{\rm GHz}$ and a typical magnetic field of $H \simeq 0.1\;{\rm Tesla}$, we find
$j \simeq 0.05$. In this adiabatic limit, the microwave field can not give 
rise to transitions between Landau levels, thus neither elastic nor inelastic ZRS theories 
can justify a strong drop in resistance of around 50\%.
The only expected effect is that of a weak heating leading to a thermal broadening of the Landau levels. 
Moreover for $\omega \tau_{tr} \gg 1$ the electric field penetrates the sample in the form of plasmon excitations.
At frequencies $\omega < \omega_c$ bulk-magnetoplasmons excitation are evanescent 
thus we expect the excitation field to be screened in the bulk of the sample \cite{volkov}.
On the contrary edge magneto-plasmon excitations are gapless and appear even at frequencies $\omega \ll \omega_c$
which may lead to an enhancement of the microwave field near the edges of the sample.
As a consequence the effect of irradiation should be confined to the sample edges.
This and recent results from \cite{toulouse} lead us to develop a model explaining the observed drop of resistance through 
the dynamics of orbits skipping along the sample edge under adiabatic microwave fields.

We first use the Landauer formula to make a connection between the four terminal resistance $R_{xx}$ 
and the drift velocities of the skipping orbits along sample edges.
This formula relates $R_{xx}$ 
to the transmission $T_n$ of the channels propagating along sample edges: $R_{xx} = \frac{h}{2 e^2 N} \sum (1 - T_n)/\sum T_n$,
where $N$ is the number of occupied Landau levels; $N \simeq 70$ at $H = 0.1\;{\rm Tesla}$ \cite{Datta}.
For this magnetic field, the typical transmission $T = 1 - N R_{xx} (2 e^2/h) \simeq 0.985$ 
is very close to unity ($R_{xx} \simeq 2.5\;{\rm \Omega}$).
Since $N$ is high in our experiments, we can make a semi-classical approximation for the transmissions: 
$T_n \simeq 1 - \frac{L}{v_g(n) \tau_n}$ where $L$ is the distance between voltage probes,
$v_g(n)$ is the group velocity of the channel which is given by the drift velocity in the semiclassical limit. 
Here $\tau_n$ is the typical time after which an electron from channel $n$ is scattered 
into the sample bulk ($\tau_n$ is however longer than $\tau_{tr}$ because
the probability of scattering back to the edge is high after a collision on an impurity \cite{buttiker}); this yields
\begin{align}
R_{xx} =  \frac{h}{2 e^2 N^2} \sum_n \frac{L}{v_g(n) \tau_n}
\label{Landauer}
\end{align}
This expression shows that orbits with low drift velocity give the main contribution to $R_{xx}$. 
We thus start by investigating the effects of microwave irradiation on 
a typical channel propagating along the 
edge with a drift velocity $v_g \ll v_F$ where $v_F$ is the Fermi velocity
(a typical trajectory is shown on Fig.~2).
The relation Eq.~(\ref{Landauer}) allows us  
to compute the resistance $R_{xx}$ from the knowledge of the drift velocities under irradiation.
This avoids the direct computation of the transmissions from a classical billiard model \cite{toulouse} 
which is numerically more expensive.

The polarization of the field is chosen along the $y$ axis, perpendicular to the edge of the sample.
This choice is related to the experimental geometry where the AC field was created by modulating the potential 
of a thin wire that was oriented parallel to the sample edge.
Another motivation is that the ratio between the perpendicular and longitudinal components 
of the electric field at the edge is given by the Hall parameter 
$\alpha = \sigma_{xy}/\sigma_{xx} \simeq 270$ at $H \simeq 0.1\;{\rm Tesla}$. 

\begin{figure}
\centerline{\includegraphics[clip=true,width=8cm]{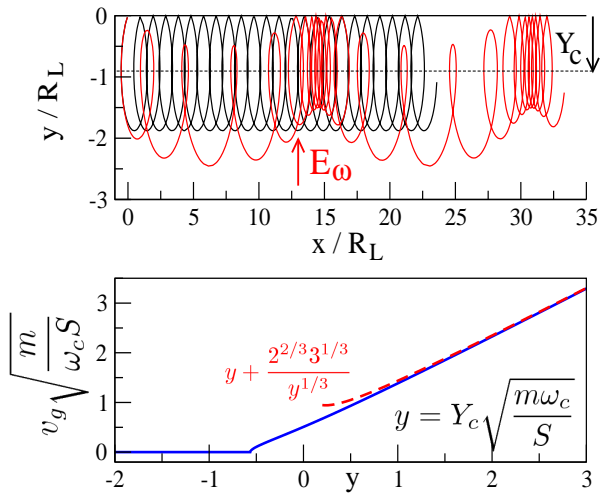}}
\caption{Top panel: Comparison between two classical trajectories propagating along the edge with the same 
initial conditions but with (red) and without (black) microwaves. The propagation is faster
in presence of driving (simulation parameters are $\omega/\omega_c = 0.1$ and $\epsilon_\omega = e E_{\omega} / (m \omega_c v_F) = 0.6$).
Bottom panel: dependence of the group velocity $v_g$ on the distance of the orbit 
guiding center to the wall $Y_c$ at fixed action $S$. The rescaled variables allow to 
obtain a functional dependence valid for all action $S$ (continuous curve),
the asymptote for high $Y_c$ is shown in dashed lines \cite{Avishai}. This high $Y_c$ limit corresponds to 
trajectories almost tangent to the wall.
}
\label{semiFig2}
\end{figure}

Two classical trajectories  with and without microwave irradiation are compared on Fig.~2, 
they start with the same initial conditions but progressively diverge due to 
the effect of microwaves. The trajectory with irradiation propagates on average faster, which
on the basis of our previous arguments, will lead to a decrease of $R_{xx}$. 
We will now show that this enhancement of drift velocity under irradiation is 
actually a general feature of edge transport and derive a simple analytical estimation 
for the increase in drift velocity. Our theoretical analysis is based 
on the conservation of the action $S$ under adiabatic driving,
which reflects the absence of transitions between Landau levels in the limit $\omega \ll \omega_c$. 
In absence of irradiation the drift velocity $v_g$ is a function of the action $S$ 
and of the position $Y_c$ of the guiding center with respect to the wall $v_g = v_g(S, Y_c)$.
The dependence on $Y_c$ is illustrated in the bottom panel of Fig.~2; the expression for 
$S$ and calculation details are given in \cite{Avishai}.
The application of a microwave irradiation induces a modulation of the position of the 
guiding center $Y_c \rightarrow Y_c + \delta Y \cos \omega t$, 
where $\delta Y = \frac{e E_{\omega}}{m \omega_c^2}$; 
$E_{\omega}$ is the amplitude of the microwave field, $\omega_c = e H/ m$ is the cyclotron frequency and $m$ is the electron mass.
Thus the time-averaged drift velocity under irradiation becomes $<v_g> = <v(S, Y_c + \delta Y \cos \omega t)>$.
The results of this averaging procedure are displayed Fig.~3 
and show the dependence of $<v_g>/v_F$ (where $v_F$ is the Fermi velocity) 
on dimensionless field $\epsilon_\omega = e E_{\omega} / (m \omega_c v_F)$ which is also the ratio 
between $\delta Y$ and the Larmor radius $R_L = v_F/\omega_c$.
It confirms the increase of the drift velocity for a large range of driving field amplitudes.
A comparison with the drift velocities extracted from direct numerical integration of the dynamics along the sample edge shows that 
the adiabatic theory gives a good quantitative prediction (see Fig.~3 inset).

\begin{figure}
\centerline{\includegraphics[clip=true,width=8cm]{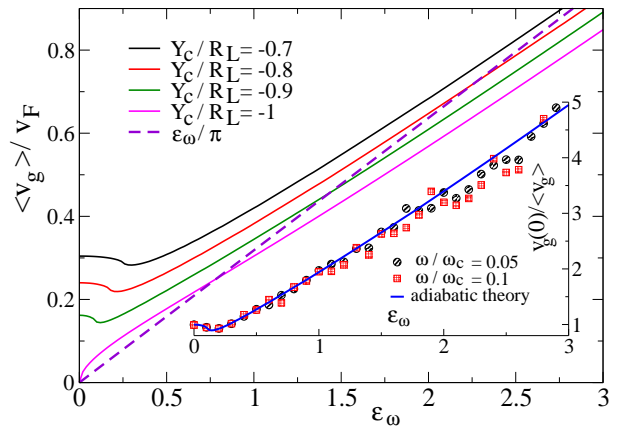}}
\caption{Time averaged drift velocity as a function of the reduced driving field $\epsilon_\omega = e E_{\omega} / (m \omega_c v_F)$
computed using the adiabatic theory for several values of the ratio between $Y_c$ and the Larmor radius $R_L = v_F/\omega_c$.
The behavior at large fields is well described by the relation $<v_g> \simeq \epsilon_\omega v_F/\pi$ represented by the dashed line.
The inset shows the good agreement between the adiabatic theory (continuous line) 
and direct numerical simulations of the classical dynamics for $Y_c = -0.9 R_L$ (symbols).
}
\label{semiFig3}
\end{figure}

The following simple argument gives a good approximation for the average drift velocity under irradiation.
The quasistatic transverse electric field $E_{\omega} \cos \omega t$ induces a drift along the wall with velocity 
$(e E_{\omega}/m \omega_c) \cos \omega t$. The equilibrium drift velocity $v_g(0)$ is enhanced when $(e E_{\omega}/m \omega_c) \cos\omega t > 0$. 
However when $(e E_{\omega}/m \omega_c) \cos\omega t + v_g(0) < 0$, the electron does not move efficiently in the direction opposite to 
its equilibrium propagation direction and the drift freezes. This behavior can be seen directly on the trajectory on Fig.~2. 
By keeping the positive contribution only, we find 
\begin{align}
<v_g> \simeq v_g(0) + \frac{e E_{\omega}}{m \omega_c \pi}
\label{SemiVgHardWall}
\end{align}
(see dashed line Fig.~3).
This expression can be also be obtained within the adiabatic formalism by retaining only the contribution from the asymptotes 
$v_g(S,Y_c) = 0$ when $Y_c \rightarrow -\infty$ and $v_g(S, Y_c) \simeq Y_c \omega_c$ for high $Y_c$ in the time averaging.
The above compact expression is compared with exact adiabatic theory on Fig.~3 and provides a satisfactory agreement.
Moreover the results of adiabatic theory are well described by straight lines 
even if the numerical coefficient derived from our heuristic argument is only approximate.
This allows to derive a simple scaling behavior for the magnetoresistance under irradiation which can be compared with our experimental data. 
For simplicity we keep the contribution of only a single typical channel propagating with drift velocity $v_g(0) \ll v_F$ in 
Eq.~(\ref{Landauer}), which leads to: 
\begin{align}
\frac{R_{xx}(0)}{R_{xx}}-1 = \frac{<v_g>}{v_g(0)}-1 \propto \frac{E_{\omega}}{\omega_c } \propto \frac{\sqrt{{\cal P}_\omega}}{H} 
\label{SemiMainEq}
\end{align}
where ${\cal P}_\omega$ is the injected microwave power and $R_{xx}(0)$ is the resistance in abscence of irradiation. Note that a scaling with the square root of microwave power
was derived for ZRS in \cite{toulouse} and observed experimentally for low temperature MIRO in \cite{mani2010}.

\begin{figure}
\centerline{\includegraphics[clip=true,width=8cm]{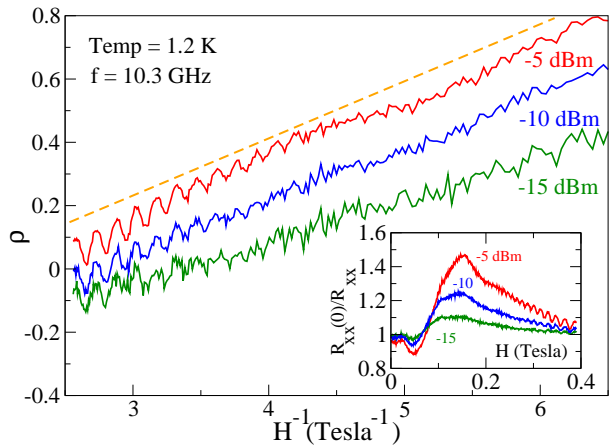}}
\caption{The solid lines show the dependence of $\rho = (P_\omega/mW)^{-1/2} (R_{xx}(0)/R_{xx}-1)$ on the inverse magnetic field for 
three power values ($-15,-10$ and $-5\;{\rm dBm}$) at frequency $10.3\;{\rm GHz}$ (we remind that $R_{xx}(0)$ is the resistance 
in abscence of irradiation). 
The curves are shifted for clarity and collapse on a single straight line (dashed curve) with slope 
independent on microwave power as predicted by Eq.~(\ref{SemiMainEq}), power was varied by an order of magnitude.
Inset shows the dependence of $R_{xx}(0)/R_{xx}$ on magnetic field for the same values of power.
}
\label{semiFig4}
\end{figure}

\begin{figure}
\centerline{\includegraphics[clip=true,width=8cm]{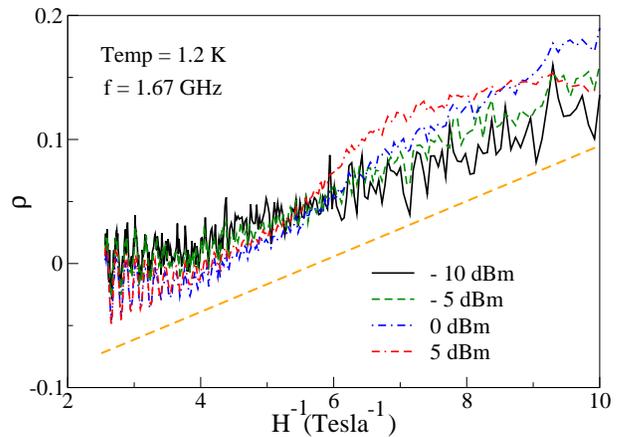}}
\caption{Dependence of $\rho = (P_\omega/mW)^{-1/2} (R_{xx}(0)/R_{xx}-1)$ on the inverse magnetic field for 
power values of $-10,-5, 0$ and $+5\;{\rm dBm}$ at frequency $f = 1.67\;{\rm GHz}$. Thus the power $P_{\omega}$ is changed by a factor of $30$. The straight dashed curve is a guide to the eye that represents the expected theoretical dependence.
}
\label{semiFig1.6GHz}
\end{figure}

The equation Eq.~\ref{SemiMainEq} predicts that the quantity $\rho = {\cal P}_\omega^{-1/2} (R_{xx}(0)/R_{xx} - 1)$ 
should vary linearly with inverse magnetic field and be independent of microwave power. 
The magnetoresistances at different microwave powers indeed collapses on a single curve according to this scaling.
This is represented on Fig.~\ref{semiFig4} for $f = 10.3 \;{\rm GHz}$, Fig.~\ref{semiFig1.6GHz} for $f = 1.67\;{\rm GHz}$, 
and a similar collapse was observed at all the other frequencies for $f \ge 1.6\;{\rm GHz}$. 
Thus this model is successful at describing the observed decrease of magnetoresistance under irradiation
in the regime of adiabatic driving $\omega \ll \omega_c$ at sufficiently strong magnetic fields 
where the $1/H$ decay is observed (see Fig.~\ref{semiFig4} inset).
At lower magnetic fields the scaling breaks down as the guiding along sample edges 
is destroyed by disorder. Our explanation relied on the enhancement of the drift velocity of skipping orbits along sample edge under
adiabatic irradiation ($\omega \ll \omega_c$) and is not suited to describe this regime. In the following, 
we will emphasize several experimental observations that appear to us relevant for 
constructing a theory valid at all magnetic fields.

\begin{figure}
\centerline{\includegraphics[clip=true,width=8cm]{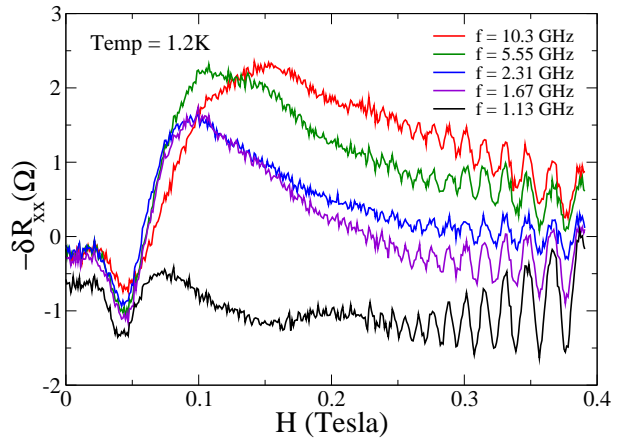}}
\caption{Variation of the magnetoresistance under illumination $-\delta R_{xx} = R_{xx}(0) - R_{xx}$  at different frequencies 
as a function of magnetic field (frequencies are listed in the legend,
and correspond to curves from top to bottom at $0.3\;{\rm Tesla}$). 
At $f = 1.13\;{\rm GHz}$ the illumination only increases $R_{xx}$, 
while for frequencies $f \ge 1.67\;{\rm GHz}$ a drop in $R_{xx}$ is observed under illumination.
In this case a scaling similar to Fig.~\ref{semiFig4} or Fig.~\ref{semiFig1.6GHz} could be constructed at all frequencies for $H \ge 0.15\;{\rm Tesla}$.
}
\label{semiFigAllFreq}
\end{figure}

The disorder potential in a high mobility 2DEG induces mainly small angle scattering.
As a consequence it is characterized by two time scales 
the elastic lifetime $\tau_{e} \simeq 20\;{\rm ps}$ which is the average time between two collisions 
and the transport lifetime $\tau_{tr} \simeq 1.1\;{\rm ns}$ which measures the time 
needed for an electron to loose memory of its momentum \cite{ando}. 
While $\tau_{tr}$ is extracted from the mobility, $\tau_{e}$ is obtained from the decay of the Shubnikov-de Haas oscillations.  
By varying microwave frequency, typical results are shown on Fig.~\ref{semiFigAllFreq}, 
we found that the decrease of resistance due to the enhancement of drift velocity
occurred only for $\omega \tau_{tr} \gg 1$ (the effect was present for $f = 1.67\;{\rm GHz}$ but absent for $f = 1.13\;{\rm GHz}$).
However the lowest magnetic field for which resistance still decreases under irradiation ($H \simeq 0.06\;{\rm Tesla}$ for data on Fig.~1)
does not seem determined by $\tau_{tr}$ but rather by $\tau_e$.
Indeed at $H \simeq 0.06 \;{\rm Tesla}$, $\omega_c \tau_{tr} \simeq 160$ while $\omega_c \tau_{e} \simeq 3$ is of the order 
of unity. At lower magnetic fields $H \le 0.06 \;{\rm Tesla}$ the resistance of the sample is enhanced under 
irradiation, this is consistent with a heating of the sample since in the explored temperature range the resistance 
increases with temperature.
Interestingly, the data represented on Fig.~\ref{semiFigAllFreq} exhibits a peak at around $H \simeq 40\;{\rm mTesla}$,
whose position does not depend on the irradiation frequency, although presently we can only speculate on the 
origin of this peak we have previously observed similar behavior when commensurability effects where present \cite{lps}.
The Larmor radius at a magnetic field of $40\;{\rm mTesla}$ is $2.4\;{\rm \mu m}$. The geometrical 
parameters of our sample are much larger than this length-scale, however the presence of  
an inhomogeneity of the 2DEG on this length-scale which could lead to the appearance of a peak at this magnetic field.
In section III. we discuss in detail, the potential effect of such an inhomogeneity on the drift velocity.

Compared to the adiabatic effect, MIRO appear only at higher frequencies in our experiments. They could be 
observed only for frequencies larger than $30\;{\rm GHz}$, suggesting that they require the absence of scattering during a 
microwave oscillation period $\omega \tau_{e} \ge 1$. However MIRO can persist
down to very low magnetic fields around $10\;{\rm mTesla}$ \cite{zudov2003}, which corresponds to $\omega_c \tau_{tr} \ge 1$. 
Therefore enhancement of guiding and MIRO/ZRS seem to appear in complementary regimes of magnetic fields and 
frequency. These observations demonstrate the importance of the two timescales $\tau_{e}$ and $\tau_{tr}$ 
for understanding phototransport in 2DEG. 

\section{II. adiabatic calculation of the group velocity}

In the previous section, we focused on the description of our experimental results and on a qualitative 
description of the enhancement of the drift velocity by a slowly varying microwave field,
a comparison was made between experiment and theory on the dependence on magnetic field and microwave power.
We will now concentrate on the derivation of the main formulas used in the calculation of the mean 
drift velocity under low frequency microwave irradiation within the adiabatic approximation.
In the Landau-Gauge the Hamiltonian for the motion of an electron along an edge in 
presence of magnetic field reads \cite{Avishai}:
\begin{align}
H = \frac{p_y^2}{2 m} + U(y) 
\end{align}
the potential $U(y)$ is created by a hard specular wall located at $y = 0$:
\begin{align}
U(y) = 
\left\{
\begin{array}{cc}
\frac{m \omega_c^2}{2} (y - Y_c)^2 & {\rm if}\; y < 0\\
\infty & {\rm if}\;y \ge 0
\end{array}
\right.
\end{align}
where $Y_c = k \hbar/(e H)$ is the position of the guiding center; and $k$ is the wavenumber in the direction 
parallel to the wall.

The oscillation period in this potential for a particle with energy $E$ and guiding center $Y_c$ reads: 
\begin{align}
T(E,Y_c) &= \frac{2}{\omega_c} {\rm Arccos}(t) \\
t &= \omega_c Y_c \sqrt{\frac{m}{2 E}} = \frac{Y_c}{R_L}
\end{align}
and integration over energy yields the expression for action:
\begin{align}
S(E, Y_c) &= \frac{2 E}{\omega_c} \sigma(t) \\
\sigma(t) &= {\rm Arccos}(t) - t \sqrt{1-t^2} 
\end{align}

In the semiclassical approximation valid for levels with number $n \gg 1$, 
the positions of the energy levels are given by the equation: 
\begin{align}
S(E_n(Y_c),Y_c) \simeq 2 \pi \hbar n 
\end{align}

Using this expression we find the value of the group velocity 
\begin{align}
v_g &= \frac{1}{\hbar}\frac{\partial E_n}{\partial k} \\
   &= -\frac{1}{m \omega_c} \frac{\partial_{Y_c} S}{\partial_E S} \\
   &= \frac{2 \sqrt{R_L^2 - Y_c^2}}{T(Y_c, E)} 
\end{align}
As expected the group velocity coincides with the drift velocity of a classical trajectory 
propagating along the sample edge with guiding a center $Y_c$.

The above equation gives an expression of $v_g$ as a function of $E, Y_c$, 
however to apply the adiabatic theory we need to evaluate $v_g$ as a function of $S, Y_c$.
For this purpose we use the following expression :  
\begin{align}
S(E, Y_c) = \frac{2 E}{\omega_c} \sigma(t) = m \omega_c Y_c^2 \sigma(t) t^{-2}
\label{eq:invert}
\end{align}
Inverting this equation should lead to an expression of $t$ as a function of $S/(m \omega_c Y_c^2)$, 
however this quantity does not depend on the sign of $Y_c$. As a result Eq.~(\ref{eq:invert}) has in 
general two solutions of opposite sign. The correct solution can then be chosen by noting 
that $t$ and $Y_c$ have the same sign. Thus we instead invert numerically the relation:
\begin{align}
\sqrt{\frac{m \omega_c}{S}} Y_c = \frac{t}{\sqrt{\sigma(t)}}
\label{eq:tvsother}
\end{align}
which gives an expression of $t$ as a function of $\sqrt{\frac{m \omega_c}{S}} Y_c$: 
\begin{align}
t = t(\sqrt{\frac{m \omega_c}{S}} Y_c)
\end{align}

As a result the rescaled group velocity $v_g \sqrt{\frac{m}{\omega_c S}}$ depends only on $\sqrt{\frac{m \omega_c}{S}} Y_c$
through the relation: 
\begin{align}
v_g \sqrt{\frac{m}{\omega_c S}} = \frac{\sqrt{t^{-2}-1}}{ {\rm Arccos}(t) } \sqrt{\frac{m \omega_c}{S}} |Y_c|
\label{eq:vgother}
\end{align}
which is displayed on Fig.~2. In the limit of high values of $\sqrt{\frac{m \omega_c}{S}} Y_c$ Eqs.~(\ref{eq:tvsother},\ref{eq:vgother}) 
can be expanded in power series to lead the asymptotic behavior shown on Fig.~2.

We now determine the conductance in presence of adiabatic microwave driving.
Let $Y_{F}(S)$ be the value of the guiding center $Y_c$ for which the Landau levels tilted by 
the presence of the wall potential intersect the Fermi level $E(S, Y_c) = E_F$ where $E_F$ is the Fermi energy.
Without microwaves the particles at the Fermi energy with action $S$ move at a mean velocity $v_g(S, Y_F(S))$.
When the low frequency irradiation is turned on, the action is not changed (adiabatic limit)
however the group velocities are modified by the presence of a the quasi-static field $E_\omega \cos \omega t$ 
\begin{align}
v_g(S, Y_F(S)) \rightarrow v_g(S, Y_F(S) + Y_\omega) - \frac{e E_\omega}{m  \omega_c} \cos \omega t
\end{align}
where $Y_\omega = E_{\omega}/(m \omega_c^2) \cos \omega t$.  
Indeed the electric field $E_\omega \cos \omega t$ changes the energy levels to : 
\begin{align}
E_n(Y_c) \rightarrow E_n(Y_c + Y_\omega) - e E_\omega \cos \omega t Y_c - \frac{e^2 E_\omega^2}{2 m \omega_c^2}
\end{align}

Averaging over the oscillations of the electric field $E_\omega \cos \omega t$, yields the expression for the 
mean drift velocity:
\begin{align}
<v_g> = <v_g(S, Y_F(S) + Y_\omega \cos \omega t)>_t 
\end{align}
This average was computed numerically leading to the results displayed on Fig.~3.

\section{III. Effect of a soft confining potential and polarization}

We will now consider the case of a soft confining potential:
\begin{align}
U(y) = 
\left\{
\begin{array}{cc}
\frac{m \omega_c^2}{2} (y - Y_c)^2 & {\rm if}\; y < 0\\
\frac{m \omega_c^2}{2} (y - Y_c)^2 + \frac{m \omega_w^2 y^2}{2} & {\rm if}\;y \ge 0
\end{array}
\right.
\end{align}
where we have introduced the frequency $\omega_w$ characterizing the stiffness of the wall.
Our motivation to study the effect of the shape of the confining potential is twofold: 
the hard wall description adopted in our model is only approximate and
a potential of this more general form can also describe smooth inhomogeneities in the 2DEG.
An indication on the presence of inhomogeneities on a length scale of a few microns, 
is given by the presence of a peak at $H \simeq 40\;{\rm mTesla}$ in the photoresistance 
data presented in Fig.~\ref{semiFigAllFreq}.

\begin{figure}
\centerline{\includegraphics[clip=true,width=8cm]{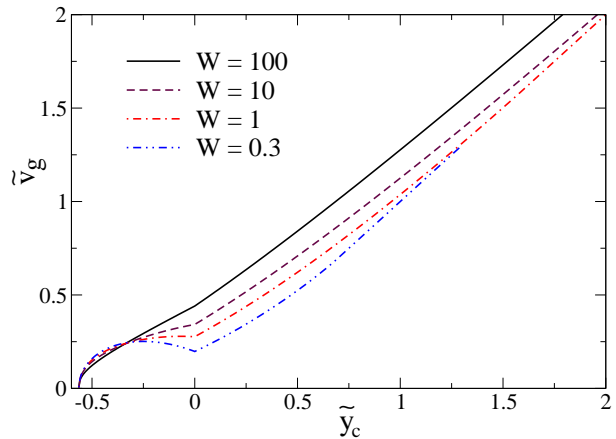}}
\caption{Dependence of the reduced drift velocity ${\tilde v}_g = v_g \sqrt{\frac{m}{\omega_c S}} \frac{\omega_c^2+\omega_w^2}{\omega_w^2}$ on the position of the guiding center ${\tilde y}_c = Y_c \sqrt{\frac{m \omega_c}{S}} $ for different 
stiffness parameters of the confinement potential $W = \omega_w^2/\omega_c^2$. 
}
\label{SemiFigSoftWall}
\end{figure}

For $y \ge 0$ the above potential can also be written as : 
\begin{align} 
U(y) &= \frac{m (\omega_c^2 + \omega_w^2)}{2} (y - Y_c')^2 + \Delta U  \\
Y_c' &= \frac{\omega_c^2 Y_c}{\omega_c^2 + \omega_w^2} \\
\Delta U &= \frac{m \omega_c^2 \omega_w^2 Y_c^2}{2 (\omega_c^2 + \omega_w^2)}
\end{align}

The Larmor radius for $y \ge 0$ is thus: 
\begin{align}
R_L' &= \sqrt{\frac{2(E - \Delta U)}{m (\omega_c^2 + \omega_w^2)}}
\end{align}
we introduce the parameter 
\begin{align}
t' &= \frac{Y_c'}{R_L'} 
= \frac{\omega_c t}{\sqrt{\omega_c^2 + \omega_w^2(1 - t^2)}}
\label{tauprime}
\end{align}
where $t = \omega_c Y_c \sqrt{\frac{m}{2 E}} = \frac{Y_c}{R_L}$.

The period of motion is then for $-1 < t < 1$ : 
\begin{align}
T(E,Y_c) = \frac{2}{\omega_c} {\rm Arccos}(t) + \frac{2}{\sqrt{\omega_c^2+\omega_w^2}} {\rm Arccos}(-t') 
\end{align}

The action is then: 
\begin{align}
S &= \frac{2 E}{\omega_c} \sigma(t) + \frac{2(E - \Delta U)}{\sqrt{\omega_c^2+\omega_w^2}} \sigma(-t')
\end{align}

The displacement along the wall during a period reads:
\begin{align}
\Delta X = -\frac{1}{m \omega_c} \partial_{Y_c} S
\end{align}
we note that the above equality can be shown to holds for any shape of the wall potential.
In the present case, however it is more convenient to compute $\Delta X$ using: 
$\Delta X = \int x' dt = \int \omega_c (Y_c - y(t)) dt$  where the time integral is taken over an oscillation period
of the motion in the $y$ direction. Performing the integration we find (when $R_L \ge |Y_c|$): 
\begin{align}
\Delta X = \frac{2 \omega_w^2}{\omega_c^2+\omega_w^2}\sqrt{R_L^2 - Y_c^2} + \frac{2 \omega_c \omega_w^2 Y_c}{(\omega_c^2+\omega_w^2)^{3/2}} {\rm Arccos}(-t')
\end{align}
As previously the drift velocity is found using: 
\begin{align}
v_g &= \frac{1}{\hbar}\frac{\partial E_n}{\partial k} = -\frac{1}{m \omega_c} \frac{\partial_{Y_c} S}{\partial_E S} = \frac{\Delta X}{T(E,Y_c)}
\end{align}

The adiabatic averaging over the slowly varying microwave field is performed using the same rescaled variables as in the previous section.
The dimensionless variable $t$ can be determined as a function of $Y_c \sqrt{\frac{m \omega_c}{S}}$ using: 
\begin{align}
Y_c \sqrt{\frac{m \omega_c}{S}} = \frac{t}{\sqrt{\sigma(t) + \frac{\sigma(-t')}{\sqrt{1+W}} - \frac{W}{(1+W)^{3/2}} t^2 \sigma(-t')}}
\end{align}
where we have introduced the parameter $W = \omega_w^2/\omega_c^2$.
The rescaled group velocity can also be expressed as a function of $t$ and $Y_c \sqrt{\frac{m \omega_c}{S}}$
\begin{align}
v_g \sqrt{\frac{m}{\omega_c S}} = \frac{W}{1+W} \frac{\sqrt{t^{-2}-1} + 
\frac{{\rm Arccos}(-t')}{\sqrt{1+W}}}{ {\rm Arccos}(t) + \frac{{\rm Arccos}(-t')}{\sqrt{1+W}} } \sqrt{\frac{m \omega_c}{S}} |Y_c|
\end{align}
This equation holds as long as $-1 \le t \le 1$, in the case where $t \le 1$ we find:  
\begin{align}
v_g \sqrt{\frac{m}{\omega_c S}} = \frac{\omega_w^2}{\omega_c^2+\omega_w^2} \sqrt{\frac{m \omega_c}{S}} Y_c
\end{align}

This suggests to compare the dependence of the rescaled drift velocity
${\tilde v}_g = v_g \sqrt{\frac{m}{\omega_c S}} \frac{\omega_c^2+\omega_w^2}{\omega_w^2}$
on the position of the guiding center ${\tilde y}_c = \sqrt{\frac{m \omega_c}{S}} Y_c$ for 
wall potentials of different stiffness.
This dependence is represented on Fig.~\ref{SemiFigSoftWall} for different values of the parameter $W = \omega_w^2 / \omega_c^2$, 
it appears that the functional dependence is actually very similar for all values of $W$.
Indeed the main difference between the curves is the apparition of a kink at $Y_c = 0$ for $W \sim 1$.

The same arguments that were employed in the derivation of Eq.~(\ref{SemiVgHardWall}) in the first section, yield the following 
approximation for the mean group velocity under irradiation : 
\begin{align}
<v_g> \simeq v_g(0) + \frac{\omega_c}{\pi} \frac{\omega_w^2}{\omega_c^2+\omega_w^2} \frac{e E_\omega}{m \omega_c^2}
\end{align}
This expression was derived for an AC electric field perpendicular to the wall.
In the case where the confining potential represents the boundaries of the sample, this assumption 
on the polarization follows from the combination of the hydrodynamic boundary condition at the edges 
$\mathbf{j} . \mathbf{n} = 0$ where the vector $\mathbf{n}$ is normal to the boundary with the relation 
$\mathbf{j} = {\hat \sigma} \mathbf{E}_\omega$ where ${\hat \sigma}$ is the mobility tensor 
(we remind that in our regime of magnetic fields $\sigma_{xy} \gg \sigma_{xx}$). 

If this potential represents inhomogeneities in the 2DEG (or smooth anharmonic components of the disorder potential),
the case of parallel polarization must be considered as well. In this case an exact analytic treatment
becomes difficult, because the dynamics can not be reduced to a one dimensional Hamiltonian anymore.
At an heuristic level one can argue, that a parallel electric field $E_{\omega,x}$ will modulate 
the position of orbit centers by $Y_{\omega} = \frac{e E_{\omega,x}}{m \omega_c \omega}$ 
which, on the basis of the results obtained for the perpendicular polarization, will change the drift velocity to:
\begin{align}
<v_g> \simeq v_0(0) + \frac{\omega_c}{\pi} \frac{\omega_w^2}{\omega_c^2+\omega_w^2} \frac{e E_{\omega,x}}{m \omega_c \omega}
\end{align}
Although this argument is only approximate, we have checked that the results are in good agreement
with numerical simulations.

The expected theoretical dependence on the magnetic field is therefore different for the two polarizations.
The experimental data is consistent with a $1/\omega_c$ dependence which appears 
for the case of a hard wall potential $\omega_w \gg \omega_c$ with a polarization 
perpendicular to the wall. However this dependence was observed only in a limited range of 
magnetic fields which does not allow to exclude a scenario where the drift velocity is enhanced 
in the bulk of the sample. In this case the electrons would drift along inhomogeneities in the 2DEG,
that can create a soft anharmonic potential. 
We stress that in both cases the drift velocity scales as $E_{\omega} \propto \sqrt{P_{\omega}}$ 
in good agreement with the observed dependence on microwave power $P_{\omega}$ (see Figs.~\ref{semiFig4},\ref{semiFig1.6GHz}). 

Recent experiments on the power dependence of MIRO also reported a $\sqrt{P_{\omega}}$ dependence.
This dependence was analyzed in term of the radiation-driven orbits model \cite{inarrea1,inarreaPRL}. 
Since the submission of our article an extension of 
this model to the low frequency limit was proposed which describes successfully some aspects 
of our experiments \cite{inarrea2}. The dependence on $\sqrt{P_{\omega}}$ appears in this model, 
because it is argued that the scattering events which give the most significant displacement 
of the electron orbits occur at a certain phase of the microwave field.
While this assumption seems phenomenologically successful the underlying physical mechanism
is not very transparent, whereas our adiabatic treatment naturally shows why only a certain 
sign of the AC electric field is effective at driving the electrons. 
Recently a multi-photon absorption theory was proposed in \cite{khodas} to describe
the $\sqrt{P_{\omega}}$ behavior. Since in this article we are dealing with the adiabatic limit $\hbar \omega \ll \hbar \omega_c$,
where photon-absorption is strongly suppressed, we believe that the explanation proposed there can not be applied
to the present case.

\section{IV. Conclusions}

In conclusion we report a strong suppression of longitudinal resistance under low frequency microwave irradiation 
in a high mobility two dimensional electron gas. This effect occurs in the regime 
where the irradiation energy $\hbar \omega$ is much smaller than the spacing between Landau levels 
$\hbar \omega_c$ and does not induce interlevel transitions. We explain our results through 
the enhancement of the mean drift velocity along sample edges by a low frequency electric field.
The theoretical analysis of this enhancement leads to a scaling relation between resistance,
power and magnetic field which is confirmed experimentally.
The adiabatic theory for the hard wall case is developed in detail in section II,
and extended in section III to the case general case of a soft confinement potential.
We have found that the described effect survives even in the limit where the characteristic frequency $\omega_w$
of the confining potential $U(y) = m \omega_w^2 y^2/2$ is comparable or smaller that the cyclotron frequency $\omega_c$.
Such a confinement potential can also appear due to inhomogeneities in the 2DEG,
and smooth components of the disorder potential.
It is therefore possible that our results could be explained by an enhancement of the mean drift velocity in the bulk.
We emphasize however that bulk magnetoplasmon modes are absent for $\omega < \omega_c$ and we expect 
the AC electric field to be stronger around the edges where edge magnetoplasmon modes are present.
As a consequence we believe that edge channels play an important role in the investigated physics.
We thank D.L. Shepelyansky for fruitful discussions and acknowledge ANR 
NanoTERRA for support.

\clearpage 


\begin{thebibliography}{99}


\bibitem{mani2002}  R.G.~Mani, J.H.~Smet, K. von Klitzing,
         V.~Narayanamurti, W.B.~Johnson, and V.~Umansky,
         Nature {\bf 420}, 646 (2002).

\bibitem{zudov2003}  M.A.~Zudov, R.R.~Du, L.N.~Pfeiffer, and K.W.~West,
        Phys. Rev. Lett. {\bf 90},  046807 (2003).

\bibitem{bykov} A.A. Bykov, A.K. Bakarov, D.R. Islamov and A.I. Toropov, 
        JETP Lett. {\bf 84}, 391 (2006).


\bibitem{denis} D. Konstantionv and K. Kono, PRL {\bf 105}, 226801 (2010) 

\bibitem{ryzhii} V.I. Ryzhii, Sov. Phys. Solid State {\bf 11}, 2078 (1970).

\bibitem{girvin} A.C. Durst, S. Sachdev, N. Read, and S.M. Girvin, Phys.
        Rev. Lett. {\bf 91}, 086803 (2003).

\bibitem{polyakov} I.A. Dmitriev, A.D. Mirlin, and D.G. Polyakov, 
        Phys. Rev. Lett. 
        {\bf 91}, 226802 (2003).

\bibitem{vavilov} M.G. Vavilov and I.L. Aleiner, 
        Phys. Rev. B {\bf 69}, 035303 (2004).



\bibitem{khodas2} I.A. Dmitriev, M. Khodas, A.D. Mirlin, D.G. Polyakov and M.G. Vavilov, Phys. Rev. B {\bf 80}, 165327 (2009)

\bibitem{mani2004} R.G. Mani, V. Narayanamurti, K. von Klitzing, J.H. Smet, 
        W.B. Johnson and V. Umansky Phys. Rev. B {\bf 69}, 161306(R) (2004);
        {\bf ibid.} {\bf 70}, 155310 (2004).

\bibitem{smetpolar} J. H. Smet, B. Gorshunov, et. al.
PRL {\bf 95}, 116804 (2005)

\bibitem{bykov2009} A.A. Bykov, JETP Lett. {\bf 89}, 575 (2009)

\bibitem{toulouse} A.D. Chepelianskii and D.L. Shepelyansky, Phys. Rev. B. {\bf 80}, 241308(R), 2009

\bibitem{mixalkov} S. A. Mikhailov, cond-mat/1011.1094

\bibitem{dorozhkin} S. I. Dorozhkin, J. H. Smet, V. Umansky, and K. von Klitzing {\bf 71}, 201306(R) (2005)

\bibitem{zudov2} M.A. Zudov, R.R. Du, J.A. Simmons and J.L. Reno, Phys. Rev. B. {\bf 64}, 201311(R) (2001)

\bibitem{Datta} S. Datta, {\it Electronic Transport in mesoscopic systems}, Cambridge Univ. Press, ISBN 0 521 59943 1, (1995)

\bibitem{volkov} V.A.Volkov, S.A.Mikhailov, in {\it Landau level spectroscopy} 
Elsevier Science Publ. B.V. North-Holland, (1991) p855

\bibitem{buttiker} M. B\"uttiker, Phys. Rev. B {\bf 38}, 9375 (1988).





\bibitem{Avishai} Y. Avishai and G. Montambaux, Eur. Phys. J. B {\bf 66}, 41 (2008); see also theoretical appendix at the end of this article.

\bibitem{lps} A. Chepelianskii and H.Bouchiat, Phys. Rev. Lett. {\bf 102}, 086810 (2009).

\bibitem{mani2010} R. G. Mani, C. Gerl, S. Schmult, W. Wegscheider and V. Umansky, Phys. Rev. B {\bf 81} 125320 (2010) 

\bibitem{ando} T. Ando, A.B. Fowler and F. Stern, Rev. Mod. Phys. 
        {\bf 54}, 437 (1982).

\bibitem{Kono2} D. Konstantionv, A.D. Chepelianskii and K. Kono, arXiv:1101.5667 



\bibitem{inarrea1} J. Inarrea, R. G. Mani, and W. Wegscheider, Phys. Rev. B {\bf 82}, 205321 (2010)

\bibitem{inarreaPRL} J. I\~narrea and G. Platero, 
        Phys. Rev. Lett. {\bf 94}, 016806 (2005).

\bibitem{inarrea2} J. Inarrea, Phys. Status Solidi (RRL) {\bf 6}, 271 (2012)

\bibitem{khodas} A.T. Hatke, M. Khodas, M.A. Zudov, L.N. Pfeiffer and K.W. West, Phys. Rev. B {\bf 84}, 241302(R) (2011)


\end{thebibliography}
\end{document}